\documentclass[a4paper]{jpconf}
\usepackage{graphicx}
\begin{document}
\title{Results from the CDMS~II  Experiment}

\author{Jodi Cooley for the CDMS~II Collaboration}

\address{Department of Physics, Southern Methodist University, Dallas, Texas 75275, USA}

\ead{cooley@physics.smu.edu}

\begin{abstract}

I report recent results and the status of the Cryogenic Dark Matter Search (CDMS~II) experiment at the Soudan Underground Laboratory in Minnesota, USA.  A blind analysis of data taken by 30 detectors between October 2006 and July 2007 found zero events consistent with WIMPs elastically scattering in our Ge detectors.  This resulted in an upper limit on the spin-independent, WIMP-nucleon cross section of 6.6 x 10$^{-44}$ cm$^{2}$ (4.6 x 10$^{-44}$ cm$^{2}$ when combined with our previous results) at the 90\% C.L. for a WIMP of mass 60 GeV/c$^{2}$.  In March 2009 data taking with CDMS~II stopped in order to install the first of 5 SuperTowers of detectors for the SuperCDMS Soudan project.  Analysis of data taken between August 2007 and March 2009 is ongoing.

\end{abstract}

\section{Introduction}
The Cryogenic Dark Matter Search (CDMS~II) experiment was designed to detect the passage of Weakly Interacting Massive Particles (WIMPs) by making simultaneous measurements of ionization and athermal phonons.  The experiment was operated at the Soudan Underground Laboratory in Soudan, Minnesota between June 2006 and March 2009.

The experiment contained a total of 30 z-sensitive ionization and phonon mediated (ZIP) detectors, 19 of which were germanium crystals ($\sim$250~g each) and 11 of which were silicon crystals ( $\sim$100~g each) stacked vertically into five towers.  The detectors were 7.6~cm diameter, 1~cm thick disks which were photolithographically patterned on their top face with aluminum fins and tungsten transition-edge sensors (TESs) to collect athermal phonon signals.  The bottom face contained two concentric electrodes, an inner and outer electrode covering $\sim$85\% and $\sim$15\% of the face respectively.  The detectors are operated at superconducting temperatures with a 3~V/cm voltage applied across.  Further details of the detectors and their operation can be found in Refs. \cite{Irwin:1995zzb} \cite{Saab:2002hf} \cite{Akerib:2005zy}.

When a particle interacts in our detectors it produces ionization and phonons.  WIMPs and neutrons will produce nuclear recoils while most backgrounds (electrons and photons) will produce electron recoils.  The ratio of ionization to phonon recoil energy, called "ionization yield", is strongly dependent on event type, providing us the ability to discriminate between nuclear and electron recoils by a factor $>10^{6}$.   

\begin{figure}[th]
\begin{center}
\includegraphics[width=28pc]{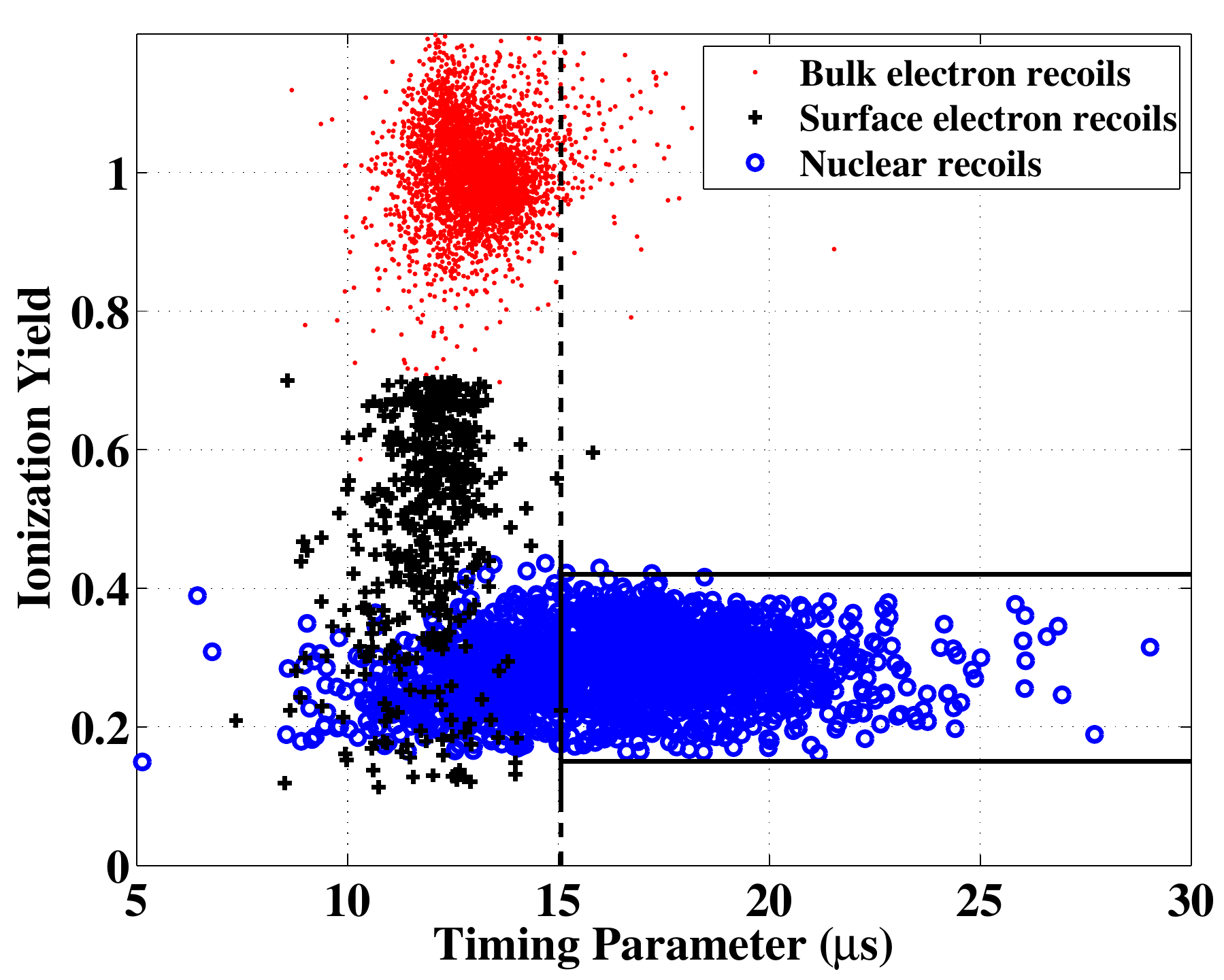}\hspace{2pc}%
\caption{\label{discr}Ionization yield versus timing parameter for calibration data in one of our Ge detectors.  Bulk electron recoils (red dots) and low yield surface events (black +) from a $^{133}$Ba source, and neutron induced nuclear recoil events (blue $\circ$) from a $^{252}$Cf source show strong dependence on ionization yield and timing parameter.  The vertical dashed line indicates minimum timing parameter for this analysis.  The boxed region shows the approximate signal region.  Taken from Ref. \cite{Ahmed:2008eu}}
\end{center}
\end{figure}

Electron recoils occurring within the first $\sim$10$\mu$m of the surface of our detectors suffer from a suppressed ionization signal which is sufficient to misclassify these events as nuclear recoils. Surface events produce phonon pulses with faster timing properties than events occurring in the bulk of the crystal.  Hence, parameterization of the phonon pulse timing properties was used to further discriminate between signal and background events.  Discrimination based on both the ionization yield and the timing parameters of the phonon pulse of events are demonstrated in Fig. \ref{discr}.

In order to reduce the overall external $\gamma$ and neutron backgrounds the detectors are shielded using layers of copper, polyetheyene, lead and an active muon veto which are described elsewhere ~\cite{Akerib:2005zy}.  To reduce the cosmogenic neutron background, the experiment is located at a depth of 2040 meters water equivalent (m.w.e.) in the Soudan Underground Laboratory, Minnesota, USA.

\section{Analysis and Results}
Here I summarize the analysis and results from data taken with all five towers operated during two periods between October 2006 and July 2007 (runs R123-R124).  A full accounting of the analysis and results can be found in Ref. \cite{Ahmed:2008eu}.  The data taken in this period resulted in a raw exposure of 397.8 kg days in our Ge detectors.  To prevent bias, we performed a blind analysis where event selection and efficiencies were calculated from calibration data and WIMP-search data outside signal region.  Events inside the signal region met the following criteria:   they were inside the detector fiducial volume, in anti-coincidence with our active muon veto, scattered in only one detector, where within the 2$\sigma$ nuclear recoil yield band and passed the phonon pulse timing cut.

The expected background due to surface interactions in this analysis was  $0.6_{-0.3}^{+0.5}(stat.)_{-0.2}^{+0.3}(syst.)$ event.  The neutron background was estimated using Monte Carlo simulations of cosmic ray muons and subsequent neutron production and transport using FLUKA, MCNPX and GEANT4.  The results were normalized to the observed veto-coincident, multiple scatter event rate.  This lead to an estimate of $<$0.1 event from neutrons in this analysis.

\begin{figure}[h]
\begin{center}
\includegraphics[width=28pc]{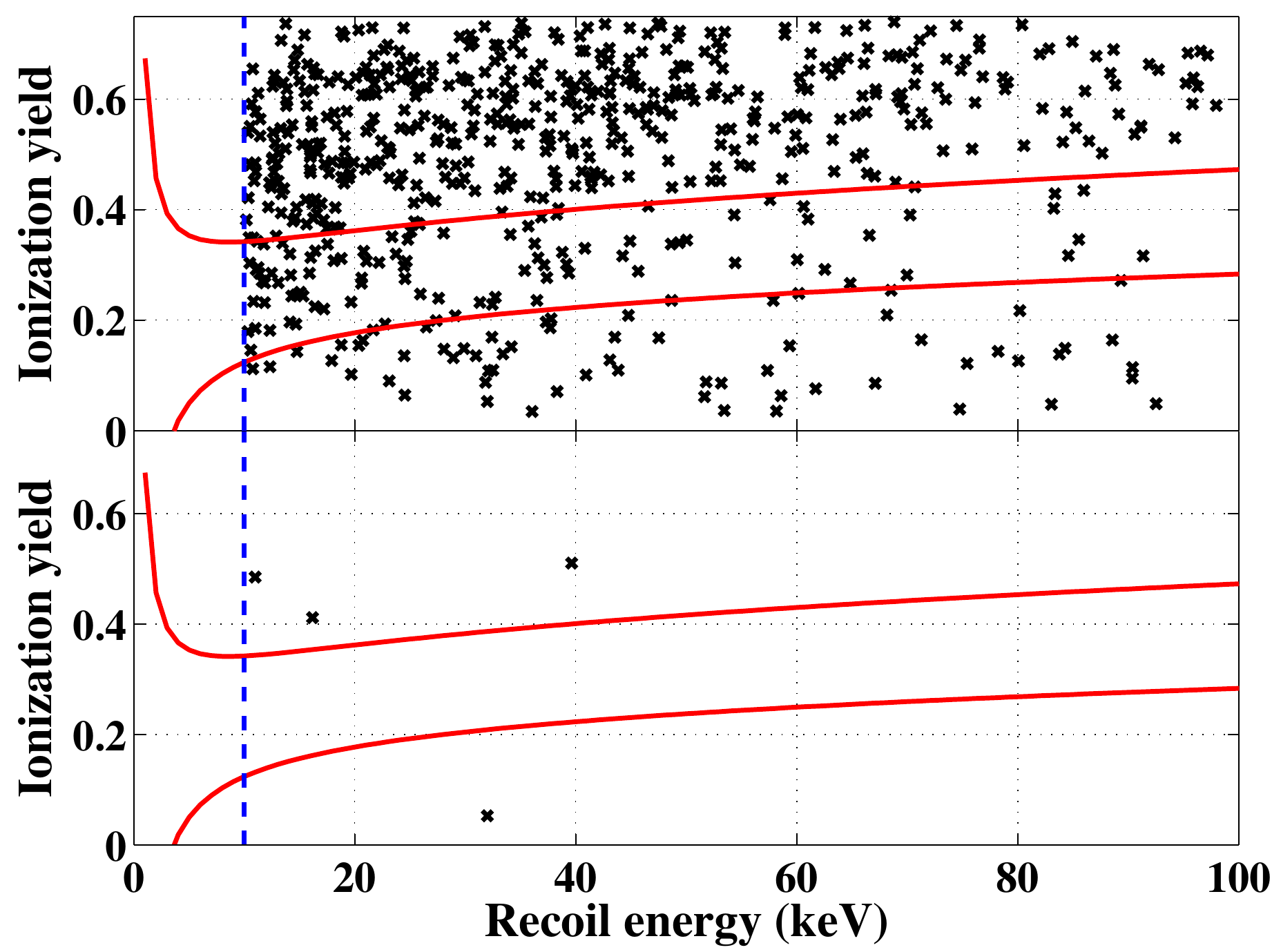}\hspace{2pc}%
\caption{\label{results}.  Top:  Ionization yield versus recoil energy for all detectors included in this analysis before timing cuts were applied.  The signal region between 10 and 100 keV (between horizontal red solid lines) was defined using neutron calibration data taken from a $^{252}$Cf source.  Bottom:  Same as top after timing cuts were applied.  No events were found in the signal region.  Taken from Ref. \cite{Ahmed:2008eu}}
\end{center}
\end{figure}

\begin{figure}[h]
\begin{center}
\includegraphics[width=28pc]{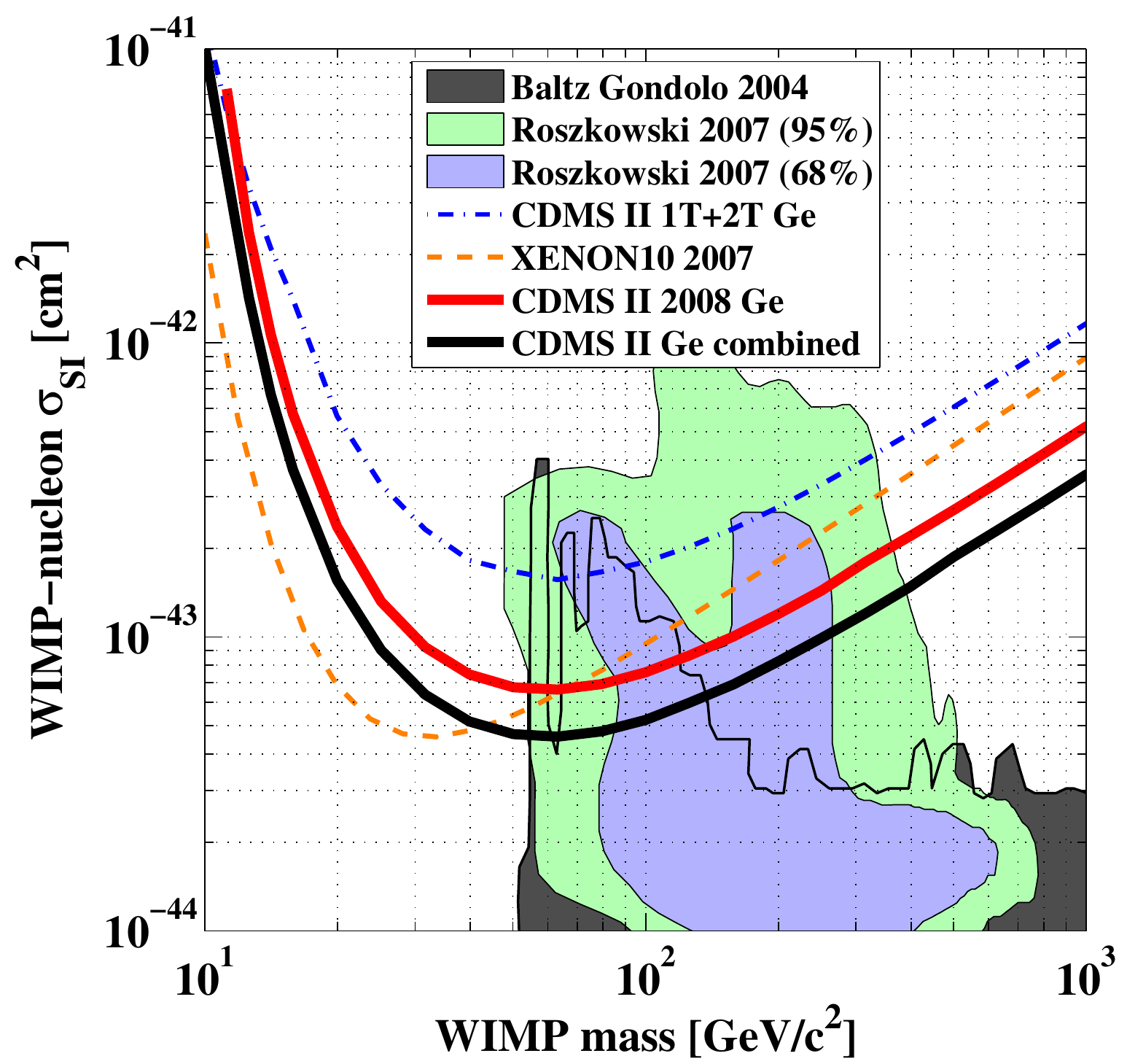}\hspace{2pc}%
\caption{\label{interpret}  WIMP-nucleon spin-Independent cross section versus WIMP mass upper limits at the 90\% C.L. from this work and other recent experiments.  The shaded region represents parameter ranges expected from supersymmetric models.  Taken from Ref. \cite{Ahmed:2008eu}}
\end{center}
\end{figure}

After applying all selection criteria, no events were observed in the WIMP-search region as shown in Fig. \ref{results}.   Under the assumptions of WIMP-nucleon spin-independent couplings and standard assumptions about our galactic halo, this resulted in a minimum upper limit of 6.6 x 10$^{-44}$ cm$^{2}$ at the 90\% confidence level for a WIMP of mass 60 GeV.  Taking into account analyses of previous data, a combined limit of 4.6 x 10$^{-44}$ cm$^{-2}$ for a WIMP of mass 60 GeV.  Figure \ref{interpret} compares the result from this work, other recent experiments and representative parameter space from SuperSymmetric models.

\section{Future}
\begin{figure}[h]
\begin{center}
\includegraphics[width=28pc]{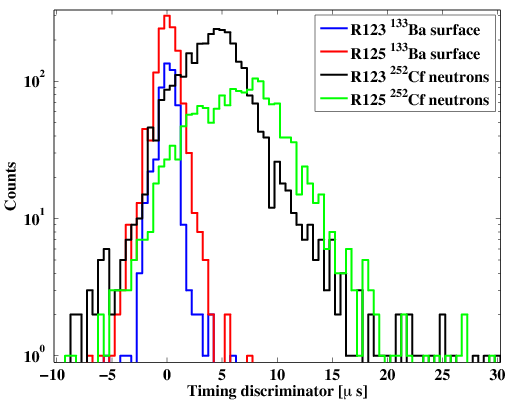}\hspace{2pc}%
\caption{\label{current} Histogram of timing parameter for calibration data taken in previous analysis and current analysis.  The timing discriminator behaves similarly for surface events (solid blue current, solid red ongoing) and neutrons (solid black current, solid green ongoing)   between the two runs.}
\end{center}
\end{figure}

In March 2009, the CDMS~II experiment stopped taking data.  The analysis of data taken between August 2007 and March 2009 (R125-R128) is on going.   The raw exposure of data taken during this time period is $\sim$1.7 times larger than the analysis reported here.  Improvements for this analysis include a faster processing package written in C++, refined timing based position correction algorithms and improved optimization of discrimination criteria based on timing of the phonon pulse.  Histograms of the phonon pulse timing parameter of calibration data indicate consistency between this data and the previous runs as shown in Fig. \ref{current}.  Results are expected to be announced in December 2009.

In order to increase the sensitivity of the the CDMS program, an increase in total exposure is needed while keeping the backgrounds under control.  The SuperCDMS Soudan project will deploy 15 kg of Ge detectors, five towers each consisting of five vertically stacked 1-inch thick detectors, in the existing Soudan setup.  For this project, the new 1-inch thick detectors with redesigned phonon sensors were developed, fabricated \cite{Brink:2009b} and tested .  The new detector thickness provides an increased mass of 2.54 over the CDMS~II detectors and the redesigned phonon sensors optimize the phonon collection area \cite{Ahmed:2009a}. The first of the five SuperTowers has been deployed and is taking data in the Soudan Underground Laboratory.

\section{References}

\medskip
\bibliographystyle{iopart-num}
\bibliography{TAUP}

\providecommand{\newblock}{}
\begin{thebibliography}{1}
\expandafter\ifx\csname url\endcsname\relax
  \def\url#1{{\tt #1}}\fi
\expandafter\ifx\csname urlprefix\endcsname\relax\def\urlprefix{URL }\fi
\providecommand{\eprint}[2][]{\url{#2}}

\bibitem{Irwin:1995zzb}
Irwin K~D, Nam S~W, Cabrera B, Chugg B and Young B~A 1995 {\em Rev. Sci.
  Instrum.\/} {\bf 66} 5322

\bibitem{Saab:2002hf}
Saab T {\em et~al.\/} 2002 {\em AIP Conf. Proc.\/} {\bf 605} 497--500

\bibitem{Akerib:2005zy}
Akerib D~S {\em et~al.\/} (CDMS) 2005 {\em Phys. Rev.\/} {\bf D72} 052009

\bibitem{Ahmed:2008eu}
Ahmed Z {\em et~al.\/} (CDMS) 2009 {\em Phys. Rev. Lett.\/} {\bf 102} 011301

\bibitem{Brink:2009b}
Brink P~L {\em et~al.\/} (SuperCDMS) {\em To be published in LTD 13, AIP Conf.
  Proc.\/}

\bibitem{Ahmed:2009a}
Ahmed Z {\em et~al.\/} (SuperCDMS) {\em To be published in LTD 13, AIP Conf.
  Proc.\/}

\end{thebibliography}

\end{document}